# Data Management Challenges in Paediatric Information Systems

Richard McClatchey

*CCS Research Centre, University of the West of England, Bristol, UK*
*Email: Richard.McClatchey@uwe.ac.uk*

*Abstract:* There is a compelling demand for the data integration and exploitation of heterogeneous biomedical information for improved clinical practice, medical research, and personalised healthcare across the EU. The area of paediatric information integration is particularly challenging since the patient's physiology changes with growth and different aspects of health being regularly monitored over extended periods of time. Paediatricians require access to heterogeneous data sets, often collected in different locations with different apparatus and over extended timescales. Using a Grid platform originally developed for physics at CERN and a novel integrated semantic data model the Health-e-Child project has developed an integrated healthcare platform for European paediatrics, providing seamless integration of traditional and emerging sources of biomedical data. The long-term goal of the project was to provide uninhibited access to universal biomedical knowledge repositories for personalised and preventive healthcare, large-scale information-based biomedical research and training, and informed policy making. The project built a Grid-enabled european network of leading clinical centres that can share and annotate paediatric data, can validate systems clinically, and diffuse clinical excellence across Europe by setting up new technologies, clinical workflows, and standards. The Health-e-Child project highlights data management challenges for the future of European paediatric healthcare and is the subject of this chapter.

## 1. Background

In recent years, there has been a tremendous increase in the volume and complexity of data available to the medical research community. To enable the use of this knowledge in clinical studies, users generally require an integrated view of medical data across a number of data sources. Clinicians, the end users of medical data analysis systems, are normally unaware of the storage structure and access mechanisms of the underlying data sources. Consequently, they require simplified mechanisms for integrating diverse heterogeneous data sources to derive knowledge about those data in order to have an holistic view of patient information and thereby to deliver personalized healthcare. Demand has risen for more holistic views of patients' health so that healthcare can be delivered at the appropriate time, by the appropriate clinician, with the appropriate means at the level of individual patients.

Medical practice as well as research is intimately dependent on information technology. From DNA sequencing to laboratory testing and epidemiological analysis, clinicians and researchers produce information, as part of their daily routine and decision making. Technology has improved dramatically the quality of these activities' results, facilitating better health-care provision and more advanced biomedical research. Nevertheless, the current state of affairs is still severely restricted with respect to the kind of information that is available to clinicians:

  a) In each case, clinicians focus their activity around a particular genre of information, e.g., genetic information or laboratory test data, therefore, obtaining a rather narrow and fragmented view of the individual patient that they are examining or the disease that they are investigating.

b) For the most part, they are confined to just using information that they themselves or, in the best case, laboratories in their immediate environment generate. For research, they do access general public data banks (e.g., GenBank), but only a limited number of such resources are actually available.

c) Especially in paediatrics, longitudinal data is usually unavailable and clinicians are forced to operate based on information generated from the current state of their patients.

d) Given this fragmented nature of the primary information available, opportunities for large-scale analysis, abstraction, and modelling are very limited as well. Hence, any secondary information and value-added knowledge that comes to the hands of the clinicians is equally restricted.

e) Face-to-face conference meetings as well as reading the literature are the only means in the hands of clinicians for exchanging experiences or obtaining second opinions on rare or unclear cases.

None of the current long-term targets of the field, e.g., personalised medical care, distributed medical teams, multidisciplinary biomedical research, etc. can be realised given the present level of technology support. The vision is to deliver a universal biomedical knowledge repository and communication conduit for the future, a common vehicle by which all paediatricians will access, analyze, evaluate, enhance, and exchange biomedical information of all forms. It will be an indispensable tool in their daily clinical practice, decision making, and research. It will be accessible at any time and from anywhere, and will offer a friendly, multi-modal, efficient, and effective interaction and exploration environment.

Clearly, any effort towards this vision requires significant change in the biomedical information management strategies of the past, with respect to functionality, operational environment, and other aspects. Contrary to current practice, the vision requires that the future paediatric systems be characterised by the following:

- **Universality of information**: they should handle "all" relevant medical applications, managing "all" forms of biomedical content. Such breadth should be realised for all dimensions of content type: biomedical abstraction (from genetics, to clinical, to epidemiological), temporal correlation (from current to longitudinal), location origin (any hospital or clinical facility), conceptual abstraction (from data to information to knowledge), and syntactic format (any data storage system, from structured database systems to free-text medical notes to images).

- **Person-centricity of information**: the systems should synthesise all information that is available about each person in a cohesive whole. This should form the basis for personalised treatment of the individual, for comparisons among different individuals, and for identifying different classes of individuals based on their biomedical information profile.

- **Universality of application**: they should ideally capture "all" aspects of "all" biomedical phenomena, diseases, and human clinical behaviours. This includes growth patterns of healthy or infected organic bodies, correlations of genotype/phenotype under several conditions of health, normal and abnormal evolution of human organs, and others.

- **Multiplicity and variety of biomedical analytics**: the systems should provide a broad collection of sophisticated analysis and modelling techniques to address the great variety of specialised needs of their applications. They should synthesise several suites of disease models, decision trees and rules, knowledge discovery and data mining algorithms, biomedical similarity measures, ontology integration mappings, and other analytics tools so that clinicians may obtain multi-perspective views of the problems of concern.

- **Person-centricity of interaction**: the primary concern of any user interaction with future paediatric systems should be the persons involved. This should be realised at three levels at least. First, the system should facilitate clinicians in identifying or generating easily all information that is pertinent to their activity and should only offer to them support for their decision making and not direct decisions. Second, it should protect the privacy of the person whose data is being accessed and manipulated. Third, it should allow biomedical information exchanges and information-based collaborations among clinicians.
- **Globality of distributed environment**: a paediatric information system should be a widely distributed system, through which biomedical information sources across the world get interconnected to exchange and integrate their contents.
- **Genericity of technology**: for economy of scale, reusability, extensibility, and maintainability, the system should be developed on top of standard, generic infrastructures that provide all common data and computation management services required. In the same spirit, all integration, search, modelling, and analysis functionality should be based on generic methods as much as possible. Any specialised functionality should be developed in a customised fashion on top of them.

It was against this background that the Health-e-Child project (hereinafter referred to as 'HeC') [1] was carried out between 2006 and 2010 in order to provide uninhibited access to universal biomedical knowledge repositories for personalised and preventive healthcare and large-scale information-based biomedical research. It focused on key instances of the above features and aimed to deliver novel, robust prototype system that was characterised by them. The emphasis of the HeC effort was on "universality of information" and its corner stone was the integration of information across biomedical abstractions, whereby all layers of biomedical data (i.e. genetic, cell, tissue, organ, individual, and population layer) would be vertically integrated to provide a unified view of a person's clinical condition (see Figure 1).

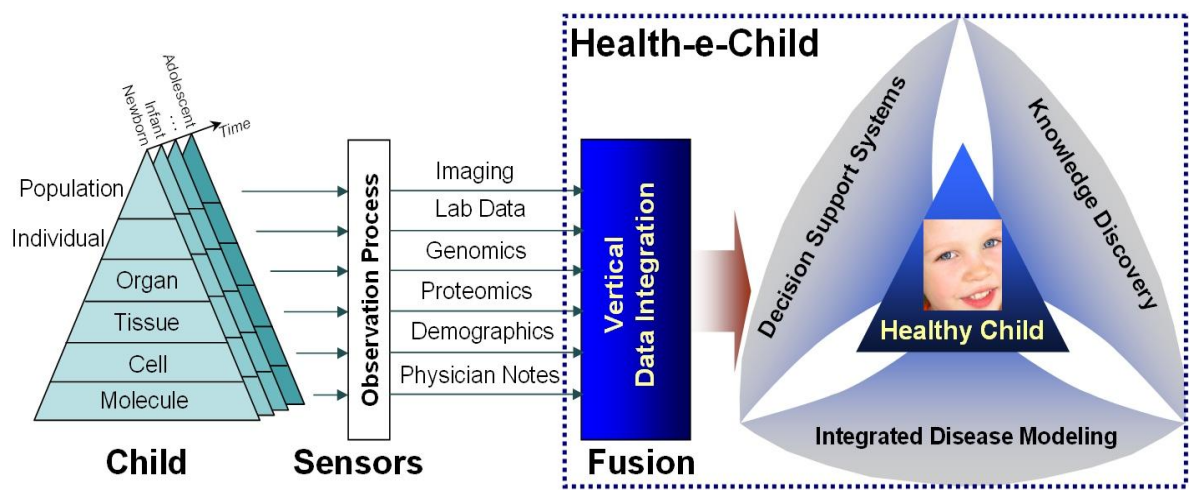

**Figure 1: The overall concept of Health-e-Child**

This drove the research and technology directions pursued with respect to all other target features of HeC, in particular, temporal and spatial information integration, information search and optimisation, disease modelling, decision support, and knowledge discovery and data mining, all operating in a distributed (Grid-based) environment.

HeC focused on paediatrics, and in particular, on carefully selected representative diseases in three different categories: paediatric heart diseases, inflammatory diseases, and brain tumours. Paediatrics adds a temporal dimension along which biomedical information changes at different speeds for the different layers of biomedical abstractions, whose vertical integration therefore faced further challenges. The particular paediatric diseases studied in HeC corresponded to largely uncharted territories with significant impact expected by any major advances in our understanding of them; they also represented a broad spectrum of technology requirements, thus ensuring genericity and broad applicability of the end result.

For economies of scale, reusability, extensibility, and maintainability, HeC was developed on top of an EGEE/gLite[1] based infrastructure that was initially deployed for physics analysis at CERN. This infrastructure provides all the common data and computation management services required by the applications. HeC is presented here as an example of how computer science (and, in particular Grid infrastructures and data models) originating from high energy physics can be adapted for use by biomedical informaticians to deliver tangible real-world benefits. The project built a Grid-enabled european network of leading clinical centres that can share and annotate paediatric data, can validate systems clinically, and diffuse clinical excellence across Europe by setting up new technologies, clinical workflows, and standards. Its work has been published in a number of journals and at several international conferences ([2], [3], [4]).

## 2. Related Work in Data Management for Paediatrics

Initiatives from which HeC benefited include the BIRN [5] project in the US, which has enabled large-scale collaborations in biomedical science by utilizing the capabilities of emerging Grid technologies. BIRN provides federated medical data, which enables a software 'fabric' for seamless and secure federation of data across the network and facilitates the collaborative use of domain tools and flexible processing/analysis frameworks for the study of Alzheimer's disease. The INFOGENMED initiative [6] has given the lead to projects in moving from genomic information to individualised healthcare and HeC built on its findings in vertical data modelling. The IHBIS project [7] proposed a broker for the integration of heterogeneous information sources in order to collect, protect and assemble information from electronic records held across distributed healthcare agencies. This philosophy is one that was also investigated in Health-e-Child. In addition, the fame-permis project [8] aimed to develop a flexible, authentication and authorisation framework to cope with security issues for a healthcare environment; aspects that were important in the delivery of the HeC prototypes. Finally the CDSS [9] project being a system that uses knowledge extracted from clinical practice to provide a classification of patients' illnesses, implemented on a Grid clearly impacted the decision support elements of Health-e-Child.

Furthermore, the MYGRID project[2] was one which indicated the benefits of an ontological approach to federated data access on the Grid in the bioscience domain. MYGRID used a web services approach as its underlying distributed systems infrastructure with an intention to migrate to Globus/OGSA based solutions at a later date. They used OWL[3] for the ontology language using description logic as opposed to the emerging WSMO (Web Service Modelling Ontology), based on first order logic [10]. WSMO is based on the Web Service Modelling Framework [11] and enables the realisation of true semantic web services, the next step in allowing Grid-based ontology mediation. Such developments are the first step in the provision of autonomous Semantic Grid systems. By adopting an ontology-based solution to unifying genetic/genomic data to patient/clinical data, HeC project took an active role in influencing the future of biomedical ontology-based Grid solutions.

---

[1] glite, A lightweight middleware for grid computing, http://glite.web.cern.ch/glite/
[2] The UK E-science project: "MYGRID – Directly Supporting the E-Scientist" http://mygrid.man.ac.uk/
[3] http://www.w3.org/TR/owl-features/ and http://www.daml.org/services/owl-s/1.0/

## 3. Specific Requirements for Paediatric Data Management

Realization of the HeC project goals (Biomedical vertical integration of data, information sharing, query processing, knowledge discovery, decision support etc.), required an infrastructure that was highly dependable and reliable. Physicians may require guarantees that the HeC system is always available and that the processes that integrate data on patients are reliable, even in the case of failures. Moreover, the infrastructure may have to allow for the transparent access to distributed data, to provide a high degree of scalability, and to efficiently schedule access to computationally intensive services by applying sophisticated load-balancing strategies. Physicians may need information immediately in order to make vital decisions. Hence, long response times due to a high system load are often not well-tolerated. Consider a case where a similarity search across a potentially large set of documents is needed. In order to support this search, feature extraction has to take place for all documents/images, nearest neighbours have to be determined, etc. All these steps require significant computing power and should not be limited to the organization where the images are stored. Rather, additional feature extraction services should be installed automatically at hosts that currently feature a low load.

The HeC project compiled a set of user requirements after interviewing the clinicians and visiting partner hospitals. These requirements were kept in mind while designing and creating the HeC Platform, described elsewhere [3]. According to these requirements, the system had to provide a distributed computing infrastructure for sharing and exchanging biomedical information and knowledge which would serve as a backbone for the HeC enabling applications. The system used gLite for virtualizing distributed computing resources and enabling secure access to sensitive medical data. The Grid middleware provided secure, coordinated and controlled access to the distributed computing resources. It facilitated the creation/removal/modification of Virtual Organizations (VOs) [12] to allow co-working between clinicians. It was possible to manage the Grid infrastructure thus allowing the 'on-the-fly' addition, removal and modification of nodes on the grid. Furthermore, according to the requirements, a "HeC Gateway" needed to be created to provide a suitable access strategy to the grid. The HeC Gateway provided access to the functionalities of the underlying grid middleware for the higher-level components of HeC, and facilitated a general API for access to the grid. The API exposed grid functionalities for the HeC services and allowed the management and monitoring of processes running on the grid and also allowed the execution of jobs on the grid, and the monitoring and analysis of the results of those jobs.

One of the major cornerstones supporting the HeC project goals was the modelling of relevant biomedical data sources. The biomedical information that is managed by HeC spans multiple vertical ranges, comes from different data sources and is possibly distributed and heterogeneous with various levels of semantic content. HeC created a set of models which facilitates the integration of all the available information that supports HeC system components, by providing access to the appropriate information between hospitals and that supports the integration across vertical levels of the medical domain.

To be able to combine all sources of data into an integrated view the model of the domain under study had to be established. Such an integrated model must provide clinicians with a coherent view of patients' health and be adaptable to changes in the models of individual sources. Some of the criteria which HeC domain models satsified include:

- capturing information specified in clinical protocols;
- supporting high-level applications such as integrated disease modelling, decision support and knowledge discovery;

- forming the basis of data management in the HeC platform and supporting the clinical queries that are expected to appear in the HeC use-cases;
- being flexible, extendible and able to evolve.

The complexity in modelling the HeC domain arose from diverse aspects:

**1. Different medical domains**, i.e. rheumatology, cardiology, neuro-oncology, radiology, genomics etc. Data collected in different departments are of different modalities, formats, structurally and semantically diverse. The models should therefore facilitate the horizontal (across different departments, hospitals, countries etc) as well as vertical (across different levels of granularity) integration. The overlapping domains have to be identified and seamlessly integrated, capturing all details from clinical protocols to establishing a common model of the patients' data.

**2. Combining anatomy/physiology models** (i.e. normal state and functioning of the body) with the disease (i.e. pathology) model. The semantics of different aspects of "normal" and "abnormal" behaviour and the correlations between different pieces of information at different levels need to be captured.

**3. Capturing functional aspects** of the sub-processes of the patient journey. The subtasks in the clinical workflows (e.g. diagnosis, treatment, follow-up etc.) may be similar in different departments sharing the same goals but the composition of these tasks as well as the strategy of realizing the tasks' goals might differ. The challenge is to identify reusable "process patterns" that can be formalized using task templates (e.g. as suggested by the CommonKADS [13] methodology) describing, for example, diagnosis or treatment procedures.

**4. Temporal dimension**. The clinical process usually follows a given time order (symptoms, study, diagnosis, treatment, follow-up, etc.), the time relates to some symptom/diagnosis and treatment concepts, time is a most important attribute for disease progression, and prognosis, and finally the temporal axis is apparent in the development of the individual. Finally, the paediatrics domain adds an additional complexity due to the fact that the child is growing and the observations in time should be aligned with the anatomical changes as well as the knowledge about how a particular disease may affect these changes. Thus, the model must cater for different time representations and aspects.

**5. Spatial (Horizontal) integration**. In clinical practice, medical data about a patient is generated at different locations, such as different departments at a hospital or at different hospitals. The patient journey from the first symptoms through diagnosis to treatment and follow-up might span multiple locations which poses several problems in maintaining the entirety of patient data. In practice, doctors often are confined to information that is produced locally; comparing similar cases or reviewing rare, complex or interesting ones is usually based on local experience. Integration along the spatial (horizontal) axis means more uniform access to possibly remote data.

**6. Vertical integration**. The information from different levels of granularity needs to be cohesively integrated to provide an integrated view of the data to the clinicians. We distinguish six levels: molecular, cellular, tissue, organ, individual and population. For example, in rheumatology the following data available at different levels can be identified:

- Molecular: genomics and proteomics data
- Cellular: results of blood tests
- Tissue: results of blood, synovial fluid tests

- Organ: bones (deformation etc.) and joints (swelling, pain, limitation of motions etc.) assessment, bones and joints measures (shape, size, bone age etc.)
- Body: lifestyle, movement assessment, damage index, rheumatology examinations
- Population: epidemiological studies based on different criteria (country of origin, sex, etc.).

The semantic link between these levels is obvious: entities are in so-called 'part-of' hierarchies, but the complexity of the human body and processes usually does not allow for drawing straightforward conclusions from parts to the whole. The HeC model had to provide access to the relevance-based, time-oriented and concept-oriented, configurable views of the data at different levels. In addition, vertically integrated data needed to be made available not only for human interpretation but also for (semi-)automated processes such as the HeC enabling tools. To facilitate this kind of integration, knowledge representation techniques were investigated and applied elsewhere in the project

**7. The linkage and usage of the external medical knowledge**. The main goal here is to establish a basis for the semantic coherence of the integrated data and provide mappings from clinical data to the external medical knowledge (e.g. biomedical ontologies and databases) to facilitate availability and accessibility of external information for querying and analysis by clinicians and applications (establishing the ground truth) as well as to make the knowledge acquired in the project available outside (sharing the knowledge).

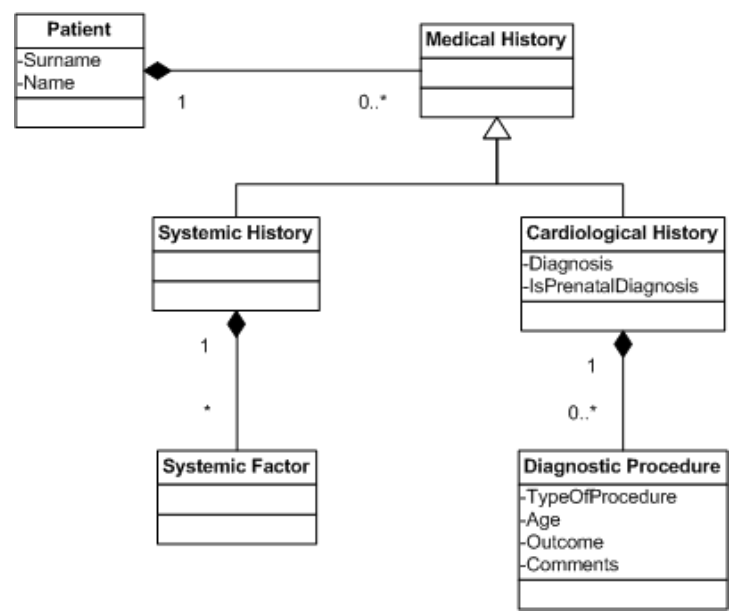

**Figure 2. An extracted fragment of the HeC data model**

As an example of the detail captured in the HeC data model Figure 2 shows a fragment of a patient's medical history as derived from a study of cardiology protocols.

**4. The HeC Integrated Data Model**

One crucial factor in the creation of integrated heterogeneous systems dealing with changing requirements is the suitability of the underlying technology to allow the evolution of the system, as studied in the CERN data management project CRISTAL [14]. A 'reflective' system utilizes an architecture where implicit system descriptions are instantiated to become explicit so-called "metadata objects" [15]. These implicit system aspects are often fundamental structures and their instantiation as metadata objects serves as the basis for handling changes and extensions to the system, making it somewhat self-describing. Metadata objects are the self-representations of the system describing how its internal elements can be accessed and manipulated. The ability to dynamically augment and re-define system specifications can result in a considerable improvement in flexibility. This leads to dynamically modifiable systems which can adapt and cope with evolving requirements. In this way we can separate the system description in terms of metadata from the particular physical representations of the data and thereby promote ease of integration of the data whilst retaining the ability for the semantics of the system to evolve.

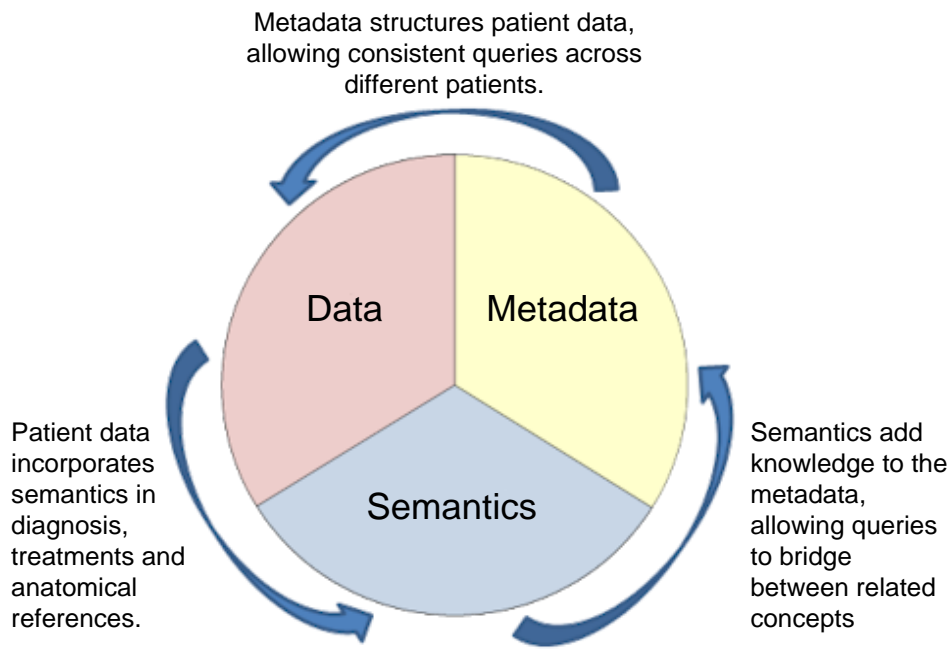

**Figure 3. A high-level overview of the HeC IDM**

The complexity which arose from the use of diverse distributed data sources in HeC and the anticipated evolution of its medical information led us to the decision to adopt a modelling approach which heavily relied on metadata. In addition the model was enhanced with a semantic layer to facilitate the semantic coherence of the integrated data and to allow linking and reuse of the external medical knowledge. The metadata reveals the structure of the underlying heterogeneous medical data allowing consistent queries across populations of patients and disease types. The semantic layer adds knowledge to this metadata thereby facilitating the resolution of queries that bridge between related concepts. It is this combination of descriptive metadata with system semantics that provided the data model with the ability to be both reactive in terms of the queries generated by user applications and to have the richness to enable integration across heterogeneous data sources. The resulting Integrated Data Model (IDM) constituted the structures for the representation of data, information and knowledge for the biomedical domain of HeC (see Figure 3).

There are many noted advantages of the use of meta-data – it can improve system interaction and data quality, it can support system and domain integration, and it can enhance system maintenance, analysis and design. In HeC the metadata structures define and describe precisely what data can be stored and how it can be accessed and this allows the evolution of the domain-specific data model without the need for re-engineering the way in which the data is actually stored or accessed. In other words the metadata describes the data structures in the system but does not address how the stored information can be interpreted or how the meaning of (a subset of) the data can be extracted to allow inference of new knowledge, potentially hidden in the data.

Developing data schema for representing structured medical information has been subject to active research and development during the last decades. HL7[4] is a standard for information exchange between different medical applications. Although the standard does convey aspects of the hospital process from the financial aspects to the handling of clinical records, HL7 is primarily a messaging standard that enables clinical applications to communicate and exchange medical data. The openEHR Foundation[5] provides a set information models and a terminology based on extendable and reusable formal models of medical concepts (archetypes) and templates allowing the domain models definition independently of terminologies used. Conceptually, the openEHR archetypes are very similar to the metadata definitions in the HeC data model. Although there is no straight correspondence between the openEHR archetypes and the HeC clinical variables types or the openEHR templates and the HeC medical event types but most certainly mappings/interfaces may be provided and the openEHR specifications.

The HeC project developed procedures for semantic query enhancement and optimization using a HeC ontology [16], described in section 5. The query answering process combined these features to: enable semantic query reformulation, enhance answer sets and improve support for approximate queries. The addition of semantic structures added flexibility and descriptive power to the integrated data model. This required the creation of a suitable knowledge representation strategy and involved the investigation and use of established and emerging ontology engineering techniques such as ontology modularization, segmentation, specification and validation.

The full detail of the HeC integrated data model is beyond the scope of this short article. Interested readers are directed to the HeC project web-site[6] for a full description of the model, on how it was implemented in the project and on how ontologies were used to assist the clinician in semantically linking data elements from diverse information sources. The following sections outline some essential aspects of the integrated data model that illuminate the challenges inherent in data management for paediatric information systems.

### 4.1 HeC IDM:Data

Conceptually all medical data of a patient can be seen as general patient information (e.g. gender, demographics, family history etc.) with a collection of atomic pieces of data (so-called clinical variables) coming from different clinical tests and procedures. The data acquisition process is organized as a set of examinations that are performed on the patient during visits where each visit gives a context/purpose for the examinations. For each patient there can be many visits (e.g. baseline and several follow-ups) at which different examinations (e.g. physical examination, imaging, laboratory test etc.) are performed to acquire different clinical variables (e.g. heart rate, blood pressure, hemoglobin level etc.). Moreover, every visit usually results in setting (or confirming) a diagnosis and/or suggesting some treatment. This information needs to be properly recorded and related to the visit.

Different examinations, diagnoses and treatments are represented as medical events i.e. something happened to the patient and was recorded at that particular point in time in the context of some medical interaction. Medical events are always associated with time which can be represented not only as instants (e.g. date of the particular examination) or intervals (e.g. drug prescription) but also relatively to some other event which might be very useful for storing uncertain or incomplete data with respect to the time (e.g. occurrence of some diseases in the past for patient's medical history).

---

[4] Health Level Seven: http://www.hl7.org/
[5] The openEHR Foundation: http://www.openehr.org/
[6] The Health-e-Child web site http://www.health-e-child.org

Clinical variables (CV) are grouped within each medical event and represent the actual clinical data as gathered by the HeC protocols. Instead of capturing CVs based on the primitive data types such as 'float', 'integer', 'string' etc. we have identified several major subclasses of CVs based on their clinical meaning and 'essence'. For instance, the aforementioned Physical Quantity pattern can represent any measurement constituted from a numeric value and a unit of measurement. The following categories of CVs have been defined within the HeC domain as important elements in integrating paediatric data (see Figure 4):

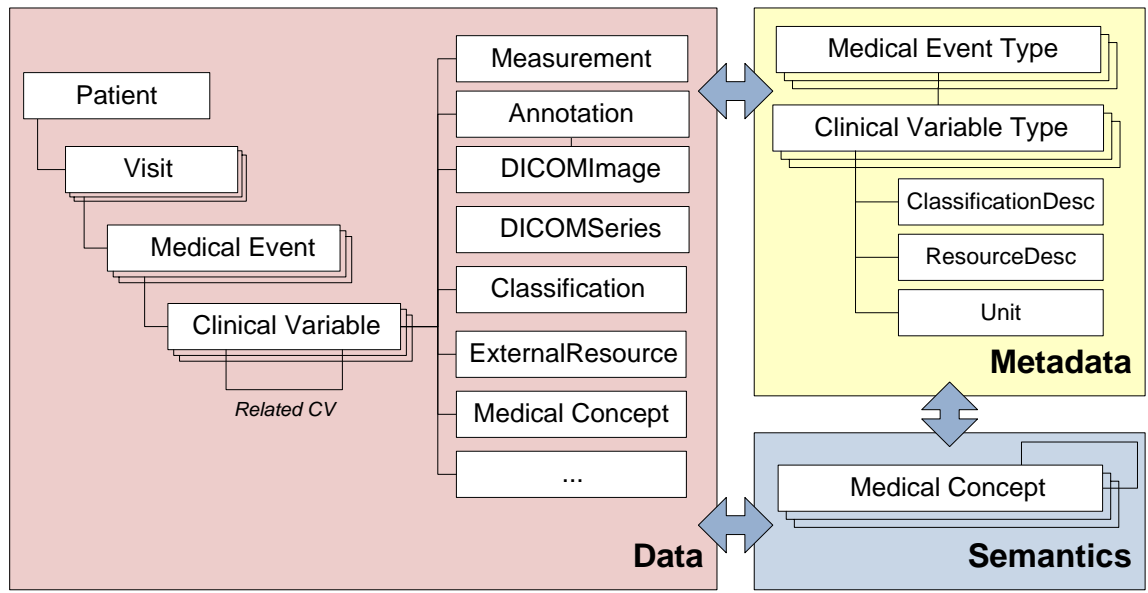

**Figure 4. Major entities of HeC IDM**

• **Measurement**: any estimation of the physical quantity (e.g. height, weight, heart rate, right ventricle (RV) volume etc.). It is important to note that each measurement has a numeric value and is associated with a unit of measurement.

• **Annotation**: any free text (e.g. comment, note, explanation etc.). Annotations can be related with any other clinical variables (of different categories) facilitating the efficient storage of any kind of the patient data with the associated annotations.

• **ObservationByClassification**: there are many clinical variables that are assessed based on some classification(s). A classification is a collection of several predefined discrete values which constrains the range of the variable values. The assessment consists of the selection of an element from this collection. For example, the assessment of the severity of RV dilation is based on the selection of one value from the predefined set of strings ("No", "Mild", "Moderate", "Severe").

• **DICOMImage**: a DICOM image can be stored in many different ways. The relevant image associated data can be extracted from the image and stored in the database to facilitate the efficient query processing without accessing the image file. DICOMData serves as a container holding the required DICOM metadata.

• **DICOMSeries**: in order to support DICOM temporal series required by the HeC applications (e.g. image registration, segmentation and 3D volume reconstruction tools) DICOMSeries class is introduced. It enables the definition of a series of DICOM images according to the specific purposes and caters for storing DICOM series from the familiar Patient-Study-Series-Image hierarchy.

- **ExternalResource**: an external resource is defined as any source of the binary data and identified by URI. In particular, a file on the Grid is considered as an external resource and the Logical File Name (LFN) that identifies the file on the Grid should be used as URI.
- **MedicalConceptInstance**: any medical event or other clinical variable can be tagged by the medical concept which is defined at the semantic level of the model (see below). For instance, the presence of a particular symptom (e.g. chest pain etc.) is captured through the instantiation of the medical concept representing the symptom as a MedicalConceptInstance object in the model.

**4.2 HeC IDM: Metadata**

There are many noted advantages of the use of meta-data [15]. They can improve system interaction and data quality, they can support system and domain integration, and they can enhance system maintenance, analysis and design. In HeC the metadata structures define and describe precisely what data can be stored and how it can be accessed for instance, medical events (MET) and clinical variables (CVT) types, measurement units, classifications and resources descriptions etc. (see Figure 4). METs and CVTs are the main organizing entities of the meta-data model. In addition to the kinds of data that can be stored in the model they define the named generic relationships between these kinds and also the grouping of these kinds according to the way the data are collected and managed at the hospitals.

Prior to storing any clinical variable in the database its description as CVTs has to be provided. CVTs provide the description of the clinical variables which represent atomic pieces of medical data. Every variable belongs to the particular category (drawing an analogy with UML, ClinicalVariable class is abstract and only its subclasses defined through the categories are instantiable) and the category of the variable is assigned through its corresponding CVT. Every CVT has a human-readable name (e.g. 'Weight', 'RV ejection fraction', 'Haemoglobin measurement' etc.) allowing the schema discovery for users and GUIs). For clarity some examples of the instantiated clinical variable and relevant structures at the metadata and semantics layers are presented in Table 1.

| SEMANTICS | METADATA | DATA |
|---|---|---|
| **Example 1.** *The measurement of Systolic LV volume is to be 30.5 mL/m$^2$* | | |
| | **CVT**: {id="SysLVVol", name="Systolic LV volume", type="Measurement"} <br> **UNIT**: {name="mL/m$^2$"} | **CV**: { type="SysLVVol", value=30.5, unit="mL/m$^2$"} |
| **Example 2.** *Patient X has severe RV dilation* | | |
| | **CVT**: {id="RVDilation", name="RV dilation", type="Classification"} <br> **Classification**: {name="Severity", items="No|Mild|Moderate|Severe"} | **CV**: { type="RVDilation", value="Severe"} |
| **Example 3.** *Patient Y has a tumour located in Cerebellum* | | |
| Cerebellum ⊑ ∃regional_part_of.Brain | **CVT**: {id="TumourLoc", name="Tumour Location", type="MedicalConceptInstance"} | **CV**: { type="TumourLoc", value="fma:Cerebellum"} |

**Table 1. Examples of data/metadata/semantics instantiation**

**4.3. HeC IDM: Semantics**

One of the aims of the HeC model was to represent the clinical data so that it could be populated and interpreted by semantic tools. The key to providing this functionality is the ability to store concepts, then specify typed relationships between these concepts, and between these concepts and the meta-data of the model. 'Concept related to concept' can then correspond directly to the 'Subject, predicate, object' declarations of RDF[7] and can be recorded at the semantics layer as MedicalConcepts with the URI (uniform resource identifier) of the knowledge source. Relating these concepts to CVTs opens up possibilities for browsing and querying software to group together relevant patient data from different patients, visits and medical events. These concepts are grouped together by types including:

- Anatomical - Body parts, organs and organ components, their relationships and characteristics relating to the three HeC clinical areas.
- Symptoms - Relating to the target diseases, including links to associated diseases.
- Diseases - By family. Differential diagnoses.
- Treatments & Drugs - Drug families, side-effects, etc.

MedicalConcepts are populated by extracting relevant fragments from the existing knowledge repositories and then need to be linked to the CVT as well as the data instances represented in the model as MedicalConceptInstance objects (again see Figure 4).

The HeC IDM metadata describes the data structures in the system but does not address how the stored information can be interpreted or how the meaning of (a subset of) the data can be extracted to allow inference of new knowledge, potentially hidden in the data. This interpretation and inference is carried out using the semantic layer which allows information to be integrated or aligned with external data sources or knowledge bases thus permitting knowledge reuse as well as making the knowledge available outside of the project [17]. Consequently, the semantics associated with the data needs to be captured to facilitate the use of integrated information.

**5. Ontologies in Paediatric Data Management**

An ontology represents a shared, agreed and detailed model of a problem domain. One advantage of the use of ontologies is their ability to resolve any semantic heterogeneity that is present within the domain data. Ontologies define links between different types of semantic knowledge. They can particularly aid in the resolution of terms for queries and other general search strategies, thus improving the search results that are presented to clinicians. The fact that ontologies are machine processable and human understandable is especially useful in this regard. There are many biomedical ontologies in existence although few, if any, support vertical integration. For example consider the Gene Ontology (GO) [18] which only defines structures regarding genes and GALEN [19] that is limited to anatomical concepts. In both cases there are no links to the other vertical levels that we have defined. The HeC project investigated the scope for reusing these ontologies, or parts thereof, which have been identified by experts in both knowledge representation and clinical matters.

Many medical ontologies do not cover the paediatrics domain adequately. For example there is a difference between the physiology of a fully grown adult and that of a child; there are also some similarities, for example they both have one heart and two lungs. Hence, it would not be sensible to reuse these ontologies in their entirety; instead HeC proposed the extraction of the relevant parts and then the integration of these into a coherent whole, thereby capturing most of the HeC domain. However integrating these ontologies into one single (upper level) ontology would not be sufficient to capture the entire HeC domain, and thus it was necessary to model the missing attributes and extend these existing ontologies to suit the project needs needs. Although there are other upper level ontologies present today, such as DOLCE [20] and SUMO [21], they were considered to be too broad to be included in the project.

---

[7] Resource Description Framework: http://www.w3.org/RDF/

The traditional ontology engineering process is an iterative process consisting of ontology modelling and ontology validation. The project chose to evaluate the different methodologies that were available for the development of a vertical domain model. These included, for example CommonKADS [13] and Diligent [22]. A methodology that deserved special consideration was proposed by Seidenberg and Rector [23] in which a strategy for modular development of ontologies was proposed, to support the re-use, maintainability and evolution of the ontology to be developed. This methodology consists of untangling the ontology into disjoint independent trees which can be recombined into an ontology using definitions and axioms to represent the relationships in an explicit fashion.

### 5.1 Ontologies and data integration

Ontologies are extensively used in data integration systems because they provide an explicit and machine-understandable conceptualization of a domain. There are several approaches to data integration as described by Wache et al. in their article [24]. In the single ontology approach, all source schemas are directly related to a shared global ontology that provides a uniform interface to the user. However, this approach requires that all sources have nearly the same view on a domain, with the same level of granularity. One example of a system using this approach is SIMS [25]. In the multiple ontology approach, each data source is described by its own (local) ontology separately. Instead of using a common ontology, local ontologies are mapped to each other. For this purpose, additional representation formalisms are necessary for defining the inter-ontology mappings.

In the hybrid ontology approach, a combination of the two preceding approaches is used. In the hybrid approach a local ontology is built for each source schema, which is not mapped to other local ontologies, but to a global shared ontology. New sources can be added with no need for modifying existing mappings. The single and hybrid approaches are appropriate for building central data integration systems, the former being more appropriate for so-called Global-as-View (GaV) systems and the latter for Local-as-View (LaV) systems. One drawback associated with the single global approach is the need for maintenance when new information sources are added to the representation. The hybrid architecture allows for greater flexibility in this regard with new sources being represented at the local level. The multiple ontology approach can be best used to construct pure peer-to-peer data integration systems, where there are no super-peers.

A mapping discovery process involves identifying similarities between ontologies in order to determine which concepts and properties represent similar notions across heterogeneous data samples in a (semi-) automatic manner. One of the major bottlenecks in generating viable integrated case data is that of mapping discovery. There exist two major approaches to mapping discovery. A top-down approach is applicable to ontologies with a well-defined goal. Ontologies usually contain a generally agreed upper-level ontology by developers of different applications; these developers can extend the upper-level ontology with application-specific terms. Examples of this approach are DOLCE [20] and SUMO [21].

A heuristics approach uses lexical structural components of definitions to find correspondences with heuristics. For example, [26] describes a set of heuristics used for the semi-automatic alignment of domain ontologies with a large central ontology. PROMPT [27] supports ontology merging, guides users through the process and suggests which classes and properties can be merged and FCA-Merge [28] supports a method for comparing ontologies that have a set of shared instances. IF-Map [29] identifies the mappings automatically by the information flow and generates a logic isomorphism [30].

Based on medical ontologies e.g. UMLS [31], GO [18] and GALEN [19] HeC investigated the mapping heuristics for integrated case data. It evaluated the relative quality of several of these mapping discovery methods for integrated case data. It concluded that a hybrid approach to ontologies using a heuristics approach to mapping discovery was suitable for its study of paediatric data integration.

### 5.2 Semantics support for DSS

Another interesting feature of ontologies is that they can aid in the creation of similarity metrics. This has already been attempted by many projects in order to gauge the similarity between genes using the GO ontology and by Resnik in [32] to gauge the similarity between different words with the WordNet Thesaurus. Investigations concluded that this technique could aid in the integration of the other sub-projects in HeC (e.g. its DSS, decision Support system) by creating a similarity metric based on the HeC ontology, hence creating a common base for the training and classification phases of the DSS. Furthermore the HeC ontology could be used within the project in the creation of ontology based training data, e.g. to classify different diseases; this can be achieved by using the rule base of the ontology within an expert system. In addition the HeC ontology could also be used to annotate different data sets such as images for easy access later, thereby creating a semantic image database.

### 5.3 Semantic query enhancement and optimization using an ontology

Ontologies as noted earlier can aid in the area of query enhancement. An example outlined here to demonstrate the principle is when an image is annotated according to the HeC ontology with the concept of a 'Jaw'. The clinician inputs a simple query into the system, presented here in natural language, stating "Give me all X-Ray images of Jaws for children with a particular disease in a specific age group" then the system will return all of the X-Rays in the database that have been annotated with the concept Jaw. However, when the system uses the power of the HeC ontology it will determine that teeth have a 'part-of' relation to a 'Jaw'. Hence the system will not only return a result set of images annotated with the concept of Jaw but will also return images annotated with the concept of teeth as well. Therefore, the clinician will be able to take advantage of a search such as this to aid in their experiments.

Query enhancement as the previous example demonstrates is important because it allows the system to provide clinicians with more targeted information. During the requirements analysis phase of the project it became clear that clinicians often struggle to create the complex queries necessary to capture all the data that they require in a study. This may cause too many or too few results to be returned thus undermining the research being undertaken.

By using the conceptual model that the HeC ontology provides we can take basic queries from users and translate them into more complex context aware searches. This reduces the amount of time taken by clinicians to locate the dataset they require which, in turn minimises the load on the system as fewer searches are necessary. Query optimisation also assists in this regard by using the HeC ontology to aid the creation of efficient data access paths by semantically altering the initial query to find a more efficient execution path within the database. Both query enhancement and optimisation are crucial in delivery of intuitive data access for clinicians whilst at the same time ensuring the scalability and overall stability of the system.

## 6. Conclusions

The HeC project ran from January 2006 and concluded in April 2010. It involved paediatric hospitals in the UK, France and Italy working closely with academics (e.g. UWE, the University of Athens, DISI-Genoa, INRIA and CERN) and with commercial partners (e.g. Siemens, Maat-G) to achieve common goals in the EC FP6 Integrated Project. The project has been able to develop complex data models to describe biomedical data from Juvenile Idiopathic Arthritis, Brain Tumours and Cardiology domains. It has also provided semantic integration of biomedical data to enable clinicians to seamlessly access data across a gLite-based Grid infrastructure based in London, Paris and Genoa. Other hospitals in Italy and the US joined in the latter stages of the project to extend this international infrastructure. Research has continued in the areas of data access and management, in the use of ontologies and in the development of suitable decision support, disease modelling and knowledge discovery applications. In order to demonstrate its findings the project developed a set of definitive paediatric clinical use-cases for the validation and testing of the applications and infrastructure during the final months of the project. It remains state-of-the-art in data integration for paediatric information management and as a valuable example of the transfer of CERN technology into healthcare.